\documentclass[multphys,vecphys]{svmult}

\usepackage{makeidx}     % allows index generation
\usepackage{graphicx}    % standard LaTeX graphics tool
\usepackage{multicol}    % used for the two-column index
\makeindex             % used for the subject index
\def\gtsim{{_>\atop{^\sim}}}
\def\ltsim{{_<\atop{^\sim}}}
\def\kms{km~s$^{-1}$}

\def\tkin{$T_{\rm kin}$}

\def\plss{Plan.\ Sp.\ Science}

\def\aap{A\&A}
\def\apj{ApJ}
\def\apjl{ApJ}
\def\hcop{HC$^{18}$O$^+$}
\def\dcop{D$^{13}$CO$^+$}
\def\hhh{H$_3^+$}
\def\hhdp{H$_2$D$^+$}
\def\ddhp{D$_2$H$^+$}
\def\ddd{D$_3^+$}
\def\nnhp{N$_2$H$^+$}
\def\nndp{N$_2$D$^+$}
\begin{document}

\title*{Chemistry and kinematics of the pre-stellar core L1544: 
        Constraints from \hhdp.}
\titlerunning{Chemistry and kinematics of L1544}
\author{Floris van der Tak\inst{1} \and Paola Caselli\inst{2}
\and Malcolm Walmsley\inst{2} \and Cecilia Ceccarelli\inst{3}
\and Aurore Bacmann\inst{4} \and Antonio Crapsi\inst{2}}
\institute{Max-Planck-Institut f\"ur Radioastronomie, Auf dem H\"ugel 69, 53111 Bonn, Germany \texttt{vdtak@mpifr-bonn.mpg.de}
\and Osservatorio Astrofisico di Arcetri, Largo E.\ Fermi 5, 50125 Firenze, Italy %\texttt{name@email.address}
\and Laboratoire d'Astrophysique de l'Observatoire de Grenoble, BP 53, 38041 Grenoble Cedex, France 
\and European Southern Observatory, Karl-Schwarzschild-Stra{\ss}e 2, 85748 Garching bei M\"unchen, Germany 
}
\authorrunning{van der Tak et al.}
\maketitle

This paper explores the sensitivity of line profiles of \hhdp, HCO$^+$
and \nnhp, observed towards the center of L~1544, to various kinematic
and chemical parameters. 
The total width of the \hhdp\ line can be matched by a static model and by
models invoking ambipolar diffusion and gravitational collapse. The
derived turbulent line width is $b$=0.15~\kms\ for the static case and
$\ltsim$0.05~\kms\ for the collapse case.
However, line profiles of \hcop\ and \nnhp\ rule out the static solution.
The double-peaked \hhdp\ line shape requires either infall speeds in
the center that are much higher than  predicted by ambipolar diffusion
models, or a shell-type distribution of \hhdp, as is the case for
HCO$^+$ and \nnhp.
%The absolute strength of the \hhdp\ line at the central position
%requires that either the central density or the adopted collisional
%rate coefficient is about twice the value adopted here.
At an offset of $\approx$20$''$ from the dust peak, the \hhdp\
abundance drops by a factor of $\approx$5.

%The shortcomings of the present model are discussed and possible
%solutions proposed.

\section{Introduction}
\label{sec:intro}

Deuterium-bearing molecules are important as probes of the very cold
phases of molecular clouds prior to star formation.  The \hhdp\ ion is
especially important as tracer of \hhh, the primary ion in dense
molecular clouds, which does not have a dipole moment and hence no
pure rotational lines.  In addition, at low temperatures
($\ltsim10$~K), \hhdp\ has the ability to channel D atoms from their
main reservoir, HD, into heavier species.  This process leads to
abundance ratios of DCO$^+$/HCO$^+$ and N$_2$D$^+$/N$_2$H$^+$ of
$\sim$10$^{-3}-10^{-1}$ observed towards dense cores, much larger than
the elemental D/H ratio of $\sim$10$^{-5}$.  Recent observations of
multiply deuterated H$_2$CO, CH$_3$OH, H$_2$S and NH$_3$ (see
Ceccarelli, this volume) suggest that under extreme conditions, a
significant fraction of D may be transferred to heavy molecules. We
wish to quantify the role of \hhdp\ in this process, and compare with
other processes such as grain surface reactions (Caselli, this
volume).

%\vfill
%\eject
  
%The figure illustrates that the light \hhdp\ molecule has a large
%energy level spacing.
The ground-state $1_{01}$--$0_{00}$ transition of para-\hhdp\ at
1370~GHz will be a prime target for GREAT on SOFIA. However, with the
upper energy level 65~K above ground, this line will only be excited
in relatively warm ($\gtsim$20~K) regions, where chemical
fractionation is ineffective. For colder sources, the
$1_{10}$--$1_{11}$ ground-state transition of ortho-\hhdp\ at 372~GHz
is more suitable, which can be observed from the ground under good
conditions.  At the low temperatures ($\ltsim$10~K) and high densities
($>10^5$~cm$^{-3}$) of pre-stellar cores, reactive collisions with
ortho--H$_2$ keep the ortho-para ratio of \hhdp\ at $\sim$1, orders of
magnitude above the LTE value \cite{pagani92,gerlich02}.

Until 2002, only two detections of \hhdp\ had been obtained, which
indicated abundances of $\sim$10$^{-11}$--10$^{-12}$ towards Class~0
objects \cite{h2d+99,stark03}.  In October 2002, we~\cite{h2d+03}
observed strong \hhdp\ emission towards the pre-stellar core L~1544,
and derived an abundance of $\sim$1$\times 10^{-9}$ in the central
$\approx$20$''$. Such a high abundance suggests that in this region,
all CNO-bearing species are depleted onto dust grains, a situation
explored in more detail by Walmsley (this volume).  Data taken in June
2003 at the CSO indicate that the same phenomenon takes place in at
least five other pre-stellar cores. More observations are scheduled
for December 2003.  These data will be presented in a forthcoming
paper. Here we investigate the line profile of \hhdp\ in L~1544, and
derive its abundance outside the central region.

\section{Kinematics}
\label{sec:kinem}

\begin{table}[b]
 \caption{Centroids and widths of the two velocity components.
Numbers in brackets denote uncertainties in units of the last decimal.}
\label{t:twopeaks}
 \centering
 \begin{tabular}{lrrr}

 \hline \noalign{\smallskip}
 \multicolumn{1}{c}{Line} & \multicolumn{1}{c}{$V_{\rm LSR}$} &
\multicolumn{1}{c}{$\Delta V_{\rm obs}$} & 
\multicolumn{1}{c}{$\Delta V_{\rm T}^a$} \cr
\multicolumn{1}{c}{} & \multicolumn{1}{c}{\kms} & 
\multicolumn{1}{c}{\kms} & \multicolumn{1}{c}{\kms} \cr
\noalign{\smallskip} \hline \noalign{\smallskip}
\hhdp\ (1$_{10}$--$1_{11}$) & 7.06(3) & 0.22(5) & 0.28--0.34 \cr
                            & 7.34(2) & 0.25(6) & \cr
\hcop\ (1--0)               & 7.04(1) & 0.18(3) & 0.10--0.12 \cr
                            & 7.28(1) & 0.23(3) & \cr 
\dcop\ (2--1)               & 7.08(2) & 0.20(4) & 0.10--0.12 \cr
                            & 7.35(4) & 0.20(8) & \cr
\nnhp\ (1--0, $F_1F$=10--11)& 7.08(1) & 0.19(1) & 0.11--0.13 \cr
                            & 7.33(1) & 0.20(2) & \cr
\noalign{\smallskip} \hline \noalign{\smallskip}
\multicolumn{4}{l}{$^a$ Thermal line width at \tkin=7 and 10 K.}
\end{tabular}
\end{table}

% Line      Area               Position           Width              Intensity
%  1  0.24409     (  0.017)   14.018 (  0.006)    0.187 (  0.013)   1.2249
%  2  0.22219     (  0.017)   14.263 (  0.008)    0.202 (  0.015)   1.0351

\begin{figure}[t]
\centering
\includegraphics[height=8cm,angle=-90]{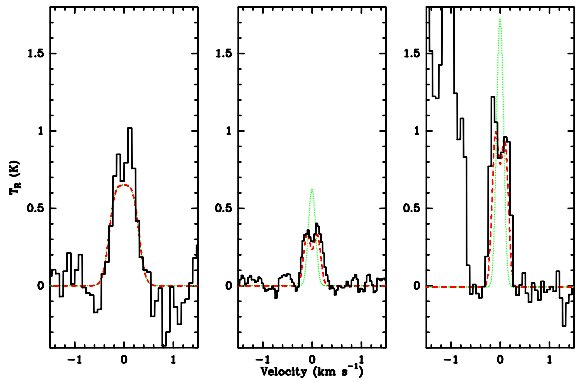}
\caption{Observed line profiles of \hhdp\ (left), \hcop\ (middle) and
  \nnhp\ (right), and synthetic profiles for the static (dotted) and
  infall (dashed) models.}
\label{fig:profi}
\end{figure}

The line profile of \hhdp\ towards L~1544 appears double-peaked,
although the signal-to-noise ratio is not high (Fig.~\ref{fig:profi}).
Fitting two Gaussians to the profile yields results very
similar to those for the profiles of \hcop, \dcop\ and \nnhp, observed
by \cite{paola02a}: two thermal components separated by
$\approx$0.26~\kms\ (Table~\ref{t:twopeaks}). We therefore investigate
whether the kinematic models that fit the HCO$^+$ and \nnhp\ data
also reproduce the \hhdp\ line profile.

%The larger width of \hhdp\ is the effect of its smaller molecular mass

\begin{figure}[t]
\centering
\includegraphics[height=6cm,angle=-90]{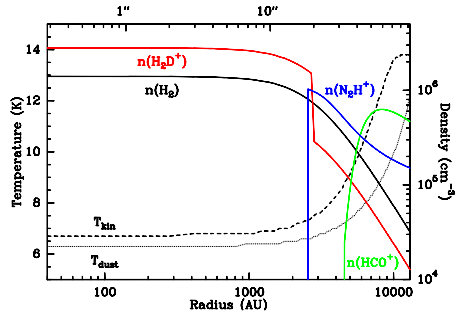}

\caption{Radial profiles of dust and gas temperature, and the
  densities of H$_2$, \hhdp\ ($\times$2$\times$10$^9$), HCO$^+$
  ($\times$10$^{12}$) and \nnhp\ ($\times$10$^{10}$) in our model of
  L~1544.}

\label{fig:model}
\end{figure}

The line profile of \hhdp\ was modeled using a Monte Carlo radiative
transfer program \cite{hst00}
\footnote{http://www.mpifr-bonn.mpg.de/staff/fvandertak/ratran/}.
Figure~\ref{fig:model} shows the adopted temperature and density structure of
L1544, taken from \cite{galli02}. See \cite{h2d+03} for details of the
excitation model; at \tkin$\ltsim$10~K, ortho-\hhdp\ is essentially a
two-level system, so that
%Only ortho-\hhdp\ was considered, using scaled radiative rates as
%collisional rate coefficients \cite{black90}. 
our results are not
sensitive to the collision rates of non-radiative transitions between
high-lying levels.
For \hhdp\ we adopt an abundance of 1$\times$10$^{-9}$ in the central
20$''$ \cite{h2d+03}. For HCO$^+$, DCO$^+$, \nnhp\ and
\nndp, we used the abundance profiles from \cite{paola02b}, Model 3,
and assumed zero abundance inside $r$=2500~AU.
We explored static models, and models with velocity fields from the
ambipolar diffusion models `t3' and `t5' of \cite{cb00} (see
\cite{paola02a} for details).  For the turbulent broadening,
Doppler parameters $b$ between 0.05 and 0.25 \kms\ were tried.
Smaller values of $b$ are overwhelmed by thermal broadening; larger
ones do not fit the data.

We find that the total width of the \hhdp\ line can be matched using either
velocity field. While the best-fit static model has $b$=0.15,
$b$=0.05 gives the best fits with the infall velocity fields. However,
the \hcop\ and \nnhp\ $J$=1-0 observations from \cite{paola02a} rule
out the static model, which does not give a double-peaked line shape.
Using the infall velocity fields, the data are matched with $b$=0.05,
consistent with the \hhdp\ results (Fig.~\ref{fig:profi}).

None of our adopted velocity fields reproduces the double-peaked
\hhdp\ line shape that the observations indicate. One possibility is
that the infall speeds of $\approx$0.1 \kms\ continue further inwards
than in the models by \cite{cb00}. Alternatively, the distribution of
\hhdp\ may have a central hole, not because of adsorption onto dust
grains (as for CO and N$_2$, the precursors of HCO$^+$ and \nnhp), but
due to conversion into \ddhp\ and \ddd.

\section{Abundance profile of \hhdp}
\label{sec:disc}

%% At the central position, the \hhdp\ models give a peak $T_R$ of
%% $\approx$0.7~K, while $\approx$0.8~K is observed.  Increasing the
%% central abundance does not increase the peak brightness because the
%% line is optically thick.  Hence \txc\ must be increased. Test models
%% indicate that the central density needs to be increased by a factor of
%% $\approx$2 from the value by \cite{galli02}. This factor is within
%%the uncertainty of the adopted collisional rate coefficient.

The abundance of \hhdp\ away from the dust peak of L~1544 was estimated to be
a factor of two lower than toward the dust peak \cite{h2d+03}.
However, this abundance may be an overestimate because some fraction of
the emission at the 20$''$ offset positions is
pickup from the central core.
We have run Monte Carlo models of the \hhdp\ emission, using the same
temperature and density structure as before, and dropping the \hhdp\
abundance at a 20$''$ radius by factors of 2--10 from its central value of
1$\times$10$^{-9}$.
The \hhdp\ intensity at the 20$''$ offset position is best matched if
the abundance drops by a factor of $\approx$5 at this radius. This
result is independent of the velocity field. Models where this factor
is 3 or 10 produce clearly worse matches to the data.

In a future paper, we will follow our results up with two-dimensional
models and an exploration of different velocity fields. We will also
model line profiles at offset positions.

% BibTeX users please use
%\bibliographystyle{}
%\bibliography{vandertakf}

% Non-BibTeX users please follow the syntax of "referenc.tex" for your own citations
%\input{referenc}
%%%%%%%%%%%%%%%%%%%%%%%% referenc.tex %%%%%%%%%%%%%%%%%%%%%%%%%%%%%%
% sample references
% "physics"
%
% Use this file as a template for your own input.
%
%%%%%%%%%%%%%%%%%%%%%%%% Springer-Verlag %%%%%%%%%%%%%%%%%%%%%%%%%%

%
% BibTeX users please use
% \bibliographystyle{}
% \bibliography{}

%%%%%%%%%%%%%%%%%%%%%%%%%%%%%%%%%%%%%%%%%%%%%%%%%%%%%%%%%%%%%%%%%%%%%%  }

%%%%%%%%%%%%%%%%%%%%%%%%%%%%%%%%%%%%%%%%%%%%%%%%%%%%%%%%%%%%%%%%%%%%%%

\printindex
\end{document}